\documentclass{article}

\usepackage{graphics}
\title{The impact of initial density profile on protoplanetary disc evolution simulation
\footnote{The present work was supported by RAS Presidium Program
18.1 "Biosphere genesis and evolution", RFBR grant 05-01-00665,
Dutch-Russian NWO-GRID project, contract NWO-RFBS 047.016.007 and
Dutch-Russian NWO-Plasma project, contract NWO-RFBS 047.016.018
and also by the Grant of Rosobrazovanie, contracts
PH$\Pi$.2.2.1.1.3653 and PH$\Pi$.2.2.1.1.1969. } }
\author{G.G.Lazareva, A.V.Snytnikov, V.A.Vshivkov}
\date{}

\begin{document}
\maketitle
\begin{abstract}
In the simulation of protoplanetary disc with a power law density
profile a disc instability is detected. The instability arises
only with a power law profile and is affected by power index.
Thus the impact of initial density profile is large within the
employed numerical model.
\end{abstract}

\section{Introduction}
Density profile is the dependence of disc surface density on
radius. As it is pointed in \cite{Raymond}, at present the true
density profiles in protoplanetary discs are unknown. Nevertheless
in most works density profile is thought to be a power law with
different index. The index should be -1.0 to correspond the
observation data \cite{DAlessio}. In simulation various indexes
in the range from -0.5 to -2.5 are being used \cite{Raymond},
\cite{Kokubo}, \cite{Lodato}.

MMSN model (Minimal Mass Solar Nebula) was proposed in
\cite{Hayashi}. The density of solid particles in this model was
obtained by imagining grinding up the planets, distributing their
mass smoothly with radius and adding up enough gas to make Solar
composition. The resulting density profile is the following:

$$
\sigma(r) = \sigma_1 \left(\frac{r}{1 \quad a.u.}\right)^{-1.5}
$$

Here $\sigma_1 = 1700\quad g/cm^{2}$, 1 a.u.$=1.5\times10^{11}
\quad m$. The ratio of gas and solid particles mass in MMSN model
is the same as in the Solar System (100:1). Unfortunately, this
model is scarcely applicable to extrasolar planetary systems as
it follows from observation of T Tauri stars \cite{DAlessio}.
Therefore the simulation of protoplanetary discs with different
initial density profiles is conducted \cite{Raymond},
\cite{Kokubo}.

The aim of the present work is to simulate the evolution of the
protoplanetary disc with different initial density profiles and
to compare the results with the known simulations of planet
system formation.

\section{Simulation}

Two density profiles were taken as initial profiles for
computational experiments: solid body profile ($\sigma_S$) and
power law profile ($\sigma_P$):
$$
\sigma_S = \sigma_1 \sqrt{1-\left(\frac{r}{R_D}\right)^2}, \qquad
r < R_D,
$$
$$
\sigma_P = \sigma_1 r^{\alpha}, \qquad \alpha = -0.5,...,-1.5.
$$
Here $R_D$ is the disc radius and the value of $\sigma_1$ is set
for the disc mass to be equal to the given value.

The computational experiments were conducted in size-less
variables in order to decrease round-off errors. The following
quantities were chosen as basic characteristic parameters for
transition to size-less variables:

--- distance from the Sun to the Earth $R_0=1.5\cdot10^{11}$ m;

--- mass of the Sun $M_\odot=2\cdot10^{30}$ kg;\

--- gravitational constant $G=6.672\cdot10^{-11}\
\mbox{Í}\cdot\mbox{m}^2/\mbox{kg}^2$.

Corresponding characteristic values of the particle velocity
($V_0$), time ($t_0$), potential ($\Phi_0$) and surface density
($\sigma_0$) are written as
$$
V_0=\sqrt{\frac{GM_\odot}{R_0}}=30\ \mbox{km/s},
$$
$$
t_0=\frac{R_0}{V_0}=5\cdot10^6\hspace{1mm} {\rm
s}=1/6\hspace{1mm} \mbox{year},
$$
$$
\Phi_0=V_0^2=\frac{GM_\odot}{R_0},
$$
$$
\sigma_0=\frac{M_\odot}{R_0^2}.
$$
In the following text all the parameters are given in size-less
units.

The ratio of central body mass, gas mass and dustsolid particles
mass was set according to the MMSN model: central body mass
$M_{\odot}=10.0$, gas mass was $M_G=1.0$ and solid particles mass
$M_P=0.01$. Both solid particles and gas were given the Keplerian
velocity $v_K$ (in any point of the disc the centrifugal force is
equal to gravitational one):
$$
\frac{\sigma v_K^2}{r}=-\frac{\partial \Phi}{\partial r}
$$
Other parameters: initial disc radius $R_D = 2.0$, radius of
computational domain $R_M = 9.0$.

The dynamics of the solid particle component of protoplanetary
disc is described by the Vlasov-Liouville kinetic equation. In the
following text dustsolid particles will be called simply
particles. To consider motion of the
 gas component the equations of gas dynamics are employed. The gravitational field
is determined by Poisson equation.

If we employ the collisionless approximation of the mean
self-con\-sis\-tent field, then Vlasov-Liouville kinetic equation
is written in the following form
$$
\frac{\partial f}{\partial t} + \vec{v}\nabla f + \vec{a}
\frac{\partial f}{\partial \vec{v}} = 0,
$$
where $f(t,\vec{r},\vec{v})$ is the time-dependent one-particle
distribution
 function along coordinates and velocities,
$\vec{a} = -\nabla\Phi + \displaystyle\frac{\vec{F}_{fr}}{m}$ is
the acceleration of unit
 mass particle, $\vec{F}_{fr}$ is the friction force between gas
 and particle
 components of the medium. Gravitational potential $\Phi$ could be
divided into two parts:
$$
\Phi = \Phi_1 + \Phi_2
$$
where $\Phi_1$ presents the potential of protostar. The second
part of of potential $\Phi_2$ is determined by the additive
distribution of the moving particles and gas. $\Phi_2$ satisfies
the Poisson equation
$$
\Delta \Phi_2 = 4\pi G \Sigma\rho
$$
In the case of a flat disc the bulk density of the mobile media
$\Sigma\rho = \rho_{part} + \rho_{gas}$ is equal to zero
($\rho_{part}$ is the particle density $\rho_{gas}$ is the gas
density). At the disc with the surface density $\sigma$ there is
a shear of the normal derivative of potential. This shear gives a
boundary condition for the normal derivative of potential
$\Phi_2$:
$$
\frac{\partial \Phi_2}{\partial z} = 2\pi G \sigma
$$
The equations of gas dynamics take the following form:
$$
\frac{\partial \rho}{\partial t} + \nabla\left(\rho\vec{v}\right)
= 0
$$
$$
\rho\left[\frac{\partial \vec{v}}{\partial t} +
\left(\vec{v}\nabla\right)\vec{v}\right] =-\nabla p + \vec{F}
$$
$$
\frac{\partial E}{\partial t} + \left(\vec{v}\nabla\right)E =
-\nabla\left(p\vec{v}\right)+Q+\left(\vec{F},\vec{v}\right)-\nabla
W
$$
where $E = \left(\varepsilon + \frac{v^2}{2}\right)$ is the
density of gas full energy, $\varepsilon =
\varepsilon\left(\rho,T\right)$ is the internal energy of gas, $p
= p\left(\rho,T\right)$ is the pressure of gas, $\vec{W}= \nabla
T$ is the heat flow, $Q$ is the increase of energy due to
chemical reactions and radiation. $\vec{F}$ is the external force
which is defined by the following expression
$$\vec{F} = \rho\nabla\Phi - k_{fr}\left(\vec{u} - \vec{v}\right)$$
here $k_{fr}$ is the coefficient of friction  between gas and
particle components of the disc, $\vec{u}$ is the particle
velocity, $\vec{v}$ is the gas velocity. In the case of the flat
disc  the form of equation remains
 the same with the only exclusion: bulk density $\rho$ is replaced with
surface density $\sigma$. In this paper we shall consider only
the flat disc
 model.

 Vlasov-Liouville equation is solved by
Particles-in-Cells method. To solve the equations of gas dynamics
Fluids-in-Cells method is employed. Poisson equation is solved by
a combination of FFT and SOR methods. A detailed description of
the code could be found in \cite{PaCT}.

In the computational experiments the cylindrical coordinate system
was used, grid size is $N_R \times N_{\varphi} \times
N_Z=300\times 256\times 100$. The experiments were conducted with
the MVS-1000M multicomputer of the Siberian Supercomputer Centre
(32 Alpha21264 processors, 833 MHz).

\section{Results}

\begin{figure}[ht]
\centerline{\includegraphics{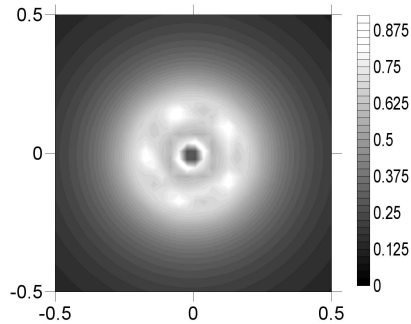}} \caption{Gas density in the
central are of the disc.}
 \label{cenDens}
\end{figure}

The most interesting result is that the disc with the massive
central body is unstable when the initial density profile
satisfies the power law. Instability here is the loss of axial
symmetry in the central are of the disc, as it is shown in figure
\ref{cenDens}: a group of dens gaseous clumps is formed around
the central body.

It is important because usually the massive central body
suppresses all the angular instabilities \cite{Bertin}.
Furthermore, this instability arises for the disc with power law
profile and does not arise for the disc with solid body profile.

In order to show the development of instability the Fourier
analysis of gas density was conducted along angular direction:
$$
\sigma(r,\varphi,t)= \sum\limits_{k=1}^{N_{\varphi}-1}
S_k(r,t)cos\left(\frac{2\pi}{N_{\varphi}}k\varphi\right)
$$
$$
S_k(r,t)= \frac{1}{N_{\varphi}}\sum\limits_{k=1}^{N_{\varphi}-1}
S_k(r,t)cos\left(\frac{2\pi}{N_{\varphi}}k\varphi\right)
$$
$$
S_{MAX}(t)= \max{S_k(r,t)}, \qquad 0 < r < R_M, \quad 1 \leq k <
N_{\varphi}
$$

\begin{figure}[ht]
\centerline{\includegraphics{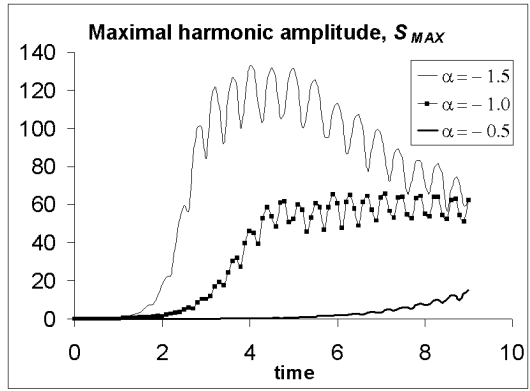}} \caption{Instability
development in the disc with power law profile for various
indexes $\alpha$.}
 \label{maxHarm}
\end{figure}

Figure \ref{maxHarm} shows maximal harmonic amplitude $S_{MAX}$
depending on time with various power indexes $\alpha$.

It should be noticed that zero harmonic ($k = 0$) is not displayed
in figure \ref{maxHarm} because this harmonic shows instabilities
that do not break axial symmetry. One can see from figure
\ref{maxHarm} that the harmonic amplitude increases sufficiently
with time.

Now let us consider the behavior of disc with various indexes
$\alpha$. Figure \ref{maxHarm} shows that with lesser values of
$\alpha$ the instability develops faster and harmonics have
greater amplitude. This fact correspond the results of
\cite{Raymond}: in their simulations discs with lower $\alpha$
formed planets earlier and the planets had greater mass.

\begin{figure}[ht]
\centerline{\includegraphics{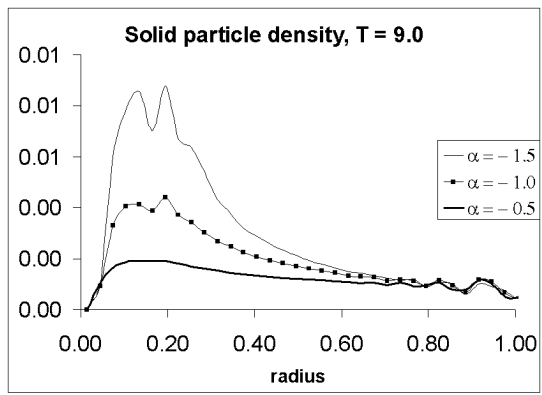}} \caption{Gas density in the
central are of the disc.}
 \label{avDens}
\end{figure}

It is also stated in \cite{Raymond} and \cite{Kokubo} that for
steeper profiles (lower values of $\alpha$), the terrestrial
planets are more massive. This result is supported by the figure
\ref{avDens} that shows average grain particle density depending
on radius.

\section{Conclusion}

It follows from the computational experiment that within the
employed model of the protoplanetary disc the impact of the
initial density profile is large. The power law profile leads to
the development of the angular instability, while the disc with
solid body profile remains stable. Moreover, the instability
develops faster with lower values of power index. The decrease of
the power index also leads to increase of amplitude of unstable
harmonics and to higher mass of gas and grain clumps.

\end{document}